# Impact of Trap Filling on Carrier Diffusion in MAPbBr$_3$ Single Crystals

N. Ganesh[1], Anaranya Ghorai[1], Shrreya Krishnamurthy[2], Suman Banerjee[1], K.L. Narasimhan[3], Satishchandra B. Ogale[2], K.S. Narayan[1]*

[1] *Chemistry and Physics of Materials Unit (CPMU), Jawaharlal Nehru Center for Advanced Scientific Research, Jakkur, Bengaluru, India – 560064*

[2] *Department of Physics and Center for Energy Sciences, Indian Institute of Science Education and Research, Pune – 411008*

[3] *Center for Nano Science and Engineering (CENSE), Indian Institute of Science Bengaluru, 560012*

## Abstract

We present experimental evidence showing that the effective carrier diffusion length ($L_d$) and lifetime ($\tau$) depend on the carrier density in MAPbBr$_3$ single crystals. Independent measurements reveal that both $L_d$ and $\tau$ decrease with an increase in photo-carrier density. Scanning photocurrent microscopy is used to extract the characteristic photocurrent $I_{ph}$ decay-length parameter, $L_d$, which is a measure of effective carrier diffusion. The $L_d$ magnitudes for electrons and holes were determined to be ~13.3 μm and ~13.8 μm respectively. A marginal increase in uniform light bias ($\leq 5 \times 10^{15}$ photons/cm$^2$) increases the modulated photocurrent magnitude and reduces the $L_d$ parameter by a factor of two and three for electrons and holes respectively, indicating that the recombination is not monomolecular. The $L_d$ variations were correlated to the features in photoluminescence lifetime studies. Analysis of lifetime variation shows intensity-dependent monomolecular and bimolecular recombination trends with recombination constants determined to be ~ 9.3 x 10$^6$ s$^{-1}$ and ~ 1.4 x 10$^{-9}$ cm$^3$s$^{-1}$ respectively. Based on the trends of $L_d$ and lifetime, it is inferred that the sub-band-gap trap recombination influences carrier transport in the low-intensity excitation regime, while bimolecular recombination and transport dominate at high intensity.

Hybrid organic-inorganic perovskites (HOIP) have demonstrated unprecedented potential for device applications such as solar cells, photodetectors, LEDs, and lasing[1-5]. Record high efficiencies, reaching 25 % [1,6,7] in HOIP single-junction solar cells can be explained due to properties such as high absorption coefficient, long diffusion lengths, decent mobility, and long-lived carrier lifetimes τ[8-10]. Long diffusion length ($L_{diff}$) in HOIP has been attributed to processes such as photon recycling, Rashba splitting, high bi-molecular recombination rates, and defect tolerance [11-15]. $L_{diff}$ estimates in the range of 100 nm – 3 mm, spanning across four orders of magnitude have been reported [8,16,17]. This difference is partly due to different techniques such as transient photoluminescence (TRPL), transient photo-voltage (TPV), transient absorption, impedance spectroscopy and transient microwave conductivity (TRMC) to determine τ [8,9,18-20]. Additionally, techniques such as space charge limited current, time of flight, Hall Effect, TRMC, and terahertz conductivity are employed to determine carrier mobility μ [8,10,17,19,21]. Determination of diffusion length ($L_{diff}$), a parameter indicative of efficient carrier transport, relies on twin measurements of τ and μ. Long $L_{diff}$ are characteristic of good quality films with large grain size. Recombination at the grain-boundaries limits the $L_{diff}$ to grain size length-scales [22]. To overcome grain size effects on $L_{diff}$ studies, sizable (2-3 mm), Methylammonium Lead Bromide ($MAPbBr_3$) single crystals (MSC) were utilized. Hybrid perovskite single crystals are evolving as suitable solar cell candidates due to low trap density, high $L_{diff}$, and high sub-bandgap absorption, reaching efficiencies of 21 % [21,23,24]. Applications based on their emission and optoelectronic properties are also emerging [25].

In the first part of the paper, we present the direct spatial estimation of effective diffusion length ($L_d$) measurements on single crystals as a function of light bias intensity using scanning photocurrent microscopy (SPCM). We then report and analyze the emission characteristics using TRPL recombination lifetime techniques. These emission studies enable understanding of the processes and the role of dc light bias on the measured $L_d$.

**Sample Details**

In this study, large MSCs (~ few mm) were prepared using the technique of inverse temperature crystallization (details in Supplemental Material) [19]. The sharp XRD peaks in **Fig. S1** is a measure of the crystal quality. SPCM studies were carried out on MSC devices. For efficient collection of photogenerated carriers, selective contact layers were deposited: poly(N,N'-bis-4-butylphenyl-N,N'-bisphenyl)benzidine (Poly-TPD) and phenyl-C71-butyric acid methyl ester ($PC_{71}BM$) with Au and Ag electrodes for hole and electron extraction respectively, schematically shown in the band diagram, **Fig. 1(a)**.

**Results**

**Scanning Photocurrent Microscopy (SPCM)**

**Fig. 1(b)** is a schematic of the set up used for SPCM (Experimental details in Supplemental Material). In this experiment, ~ 3 μm spot size, (**Fig. S2**) intensity-modulated 405 nm (~ 27 mW/cm$^2$) laser beam which is incident normally on MSC, is translated laterally between the two contacts. The modulated short-circuit photocurrent $I_{ph}(x)$ shown in **Fig. 1(c)** is measured using a lock-in amplifier at different positions of the incident beam between the two electrodes. The maxima in $I_{ph}(x)$ corresponds to the electrode-positions at either end. The contact resistance at the electrodes is estimated and verified to be negligible due to charge-selective interfacial layers, with the built-in voltage ($V_{bi}$) falling entirely along the inter-electrode length. In the present case of undoped sizable MSC (d ~ 1-2 mm, **inset Fig. 1(c)**), the built-in electric-field, $E_{bi} = V_{bi}/d$ is negligible (< 6 V/cm) resulting in a dominant diffusion contribution to the observed $I_{ph}$.

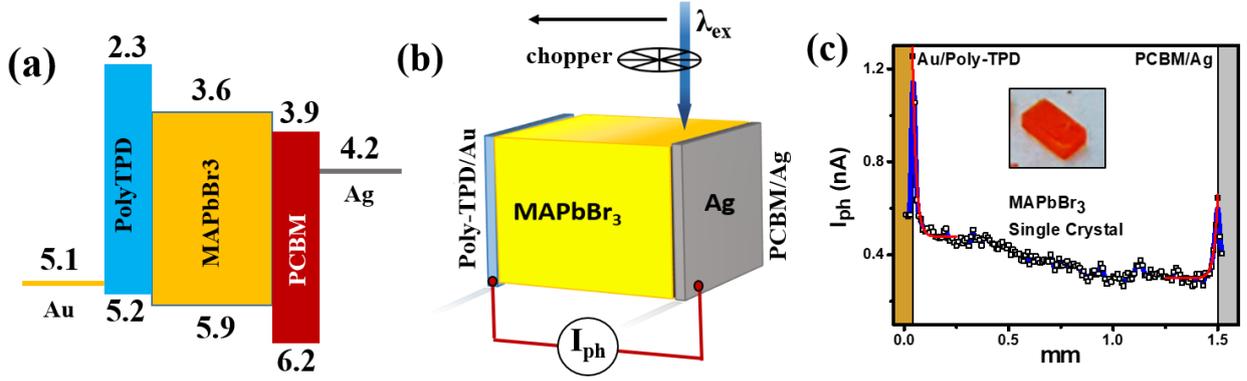

**Fig. 1: (a)** Band alignment diagram depicting MSC-device used in SPCM experiments. **(b)** Schematic of Scanning Photocurrent Microscopy (SPCM) used for effective diffusion length measurement. The incident chopped probe-beam position is varied by translating the sample. The lock-in profile $I_{ph}(x)$ is utilized to estimate $L_d$. **(c)** $I_{ph}(x)$ shows $I_{ph}$-peak at the interface. Single exponential fit to the $I_{ph}$ decay profile corresponds to the $L_d$ of the corresponding carrier. The **inset** shows the image of the bulk single crystal.

The determination of effective diffusion length, $L_d$ from the $I_{ph}(x)$ measurement can be arrived at by solving the drift-diffusion equation. In the simplified case of single-carrier 1-D transport, minority excess carrier, $\delta n$ is given as [26]:

$$\frac{\partial \delta n}{\partial t} = D \frac{\partial^2 \delta n}{\partial x^2} + \mu E \frac{\partial \delta n}{\partial x} + G - \frac{\delta n}{\tau} \qquad (1)$$

The generation rate G is given by delta function, $G = G_0 \cdot \delta(x-x_0)$ where $x = x_0$ is the position of illumination. Under conditions of steady-state illumination ($d\delta n/dt = 0$), negligible electric field and $x \neq x_0$, the diffusion current can be expressed as (derivation in Section 11, Supplemental Material):

$$I_{ph}(x) = I_0 \, exp\left(-\frac{x}{L_d}\right) \qquad (2)$$

where the effective diffusion length, $L_d$ corresponds to the $I_{ph}(x)$ decay. It must be noted that effective diffusion length, $L_d$ is different from minority carrier diffusion length, $L_{diff}$. $L_{diff}$ is the microscopic quantity related to the diffusion coefficient ($D$) and expressed as $(D\tau)^{1/2}$, where $\tau$ is the carrier lifetime. In disordered or amorphous systems, owing to the dispersive nature of transport, $L_d$ and $L_{diff}$ can differ appreciably [27,28]. However, $L_d$ has been shown to be a representative of the long-lived carriers corresponding to states in the tail distribution, in organic and polymeric semiconductors [28,29]. The $L_d$ parameter in the present case of HOIP single crystals, is expected to be a closer representation of $L_{diff}$ owing to dominant band transport [30]. In the present case of SPCM on MSC devices, it has to be noted that the $L_d$ parameter is obtained upon fitting over the entire decay range and accounts for additional factors such as finite beam spot, 3D carrier diffusion and ambipolar transport away from the electrodes.

The $I_{ph}(x)$ response in **Fig. 1(b)** can be qualitatively described as follows: Photo-excitation results in a point spread of electrons and holes via 3D diffusion – a fraction of which diffuse towards the electrodes. When the excitation is at a distance equivalent to many diffusion-lengths from the contacts, the diffusing photo-carriers get (i) trapped in shallow states and also (ii) recombine, resulting in a low-current. When the excitation beam is within $L_d$ from the electrode, the current increases exponentially with the probe distance (**Eq. (2)**) enabling one of the carriers to be extracted leaving the other carrier in the perovskite to transit to the other electrode. The extraction of one carrier and the long lifetime associated with the displacement of the counter carrier, renders the extracted carrier as minority, with excitation close ($x \leq L_d$) to the extraction electrode. Transient $I_{ph}$ measurements (**Fig. S3**), indicating longer $\tau_{transit}$ for excitation near the electrodes compared to a distant region, confirms this viewpoint. Previous reports of SPCM have determined the "minority carrier diffusion length" using doped crystals and charge selective Schottky barrier [17,31]. In the present case of MSCs, $L_d$ corresponds to the selectivity of the extraction layer in undoped single crystal devices.

Using **Eq. (2)**, $L_d$ for electrons and holes was determined to be 13.3 ± 0.6 μm and 13.8 ± 0.5 μm respectively. Intensity dependent studies reveal that the carrier $L_{diff}$ varies with excitation density similar to TPV studies, where $L_{diff}$ ~ 3 mm, at low-intensity excitation [8,10]. These studies of $L_{diff}$ determination relied on indirect estimation. To investigate the effect of intensity dependence on directly measured spatial $L_d$, intensity-dependent SPCM studies are carried out by (i) varying the intensity of the probe beam (ii) superposing a uniform background illumination or light-bias using a 390 nm LED with the probe beam. (The optical absorption depth (~ 125 nm) of the 390 nm light is similar to the 405 nm probe beam [32].)

**Fig. S4(a)** shows $I_{ph}(x)$ upon probe-intensity variation in the SPCM measurement. The excess generated probe-photo carriers, $\delta n_{probe}$ diffuse away from the point of generation and decays as $\delta n(x) = \delta n_{probe}(0) \cdot \exp\left(\frac{-x}{L_{diff}}\right)$. The corresponding $L_d$ values in **Fig. S4(b)** as a function of probe beam intensity shows that $L_d$ is independent of probe beam intensity. The invariance of $L_d$ can be understood since the photo-carriers decay to background carrier concentration within ~ $3L_{diff}$. The addition of light bias, on the other hand, maintains uniform charge generation across the transport length of the probe-carrier. It was ensured that the effects of light-induced halide redistribution and phase segregation are minimal by maintaining the MSC-device in dark under short-circuit conditions after each measurement, under a positive pressure of the inert atmosphere [33,34]. **Fig. S5** provides evidence of the sample "self-recovering" as the ions equilibrate, confirmed by the increase in the magnitude of $I_{ph}(x)$. It must be noted that we have not carried any measurements under applied bias, where one expects a sizeable ionic contribution.

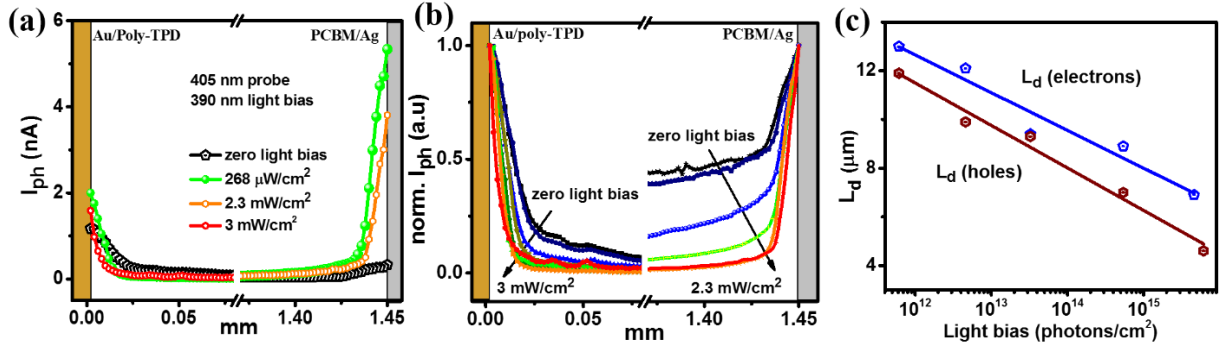

**Fig. 2:** (a) $I_{ph}(x)$ variation on both end of the MSC device with a 405 nm probe beam and 390 nm light bias. (b) Normalized $I_{ph}(x)$ profiles indicating sharp decay at higher light bias (c) $L_d$ extracted with the exponential fitting of $I_{ph}(x)$-decay as a function of light bias.

**Fig. 2(a)** depicts the $I_{ph}(x)$ profile at different dc light bias (additional plots and details in **Fig. S6**). The $I_{ph}$ magnitude increases with the light bias as indicated in $I_{ph}(x)$ maxima at the electrodes. **Fig. 2(b)** shows the normalized $I_{ph}(x)$ decay at either end of the MSC-electrode interface. The steep decay of $I_{ph}$ away from the electrode is evident under high light bias conditions. The $L_d$ parameter is extracted by fitting the decay profile to **Eq. (2)**. The estimated $L_d$ (values of $L_d$ given in **Table S1**) as a function of light bias-photon flux, presented in **Fig. 2(c)** reveals the decrease of $L_d$ with dc light-bias intensity. The extracted $L_d$ parameter decreases by a factor of 3 for holes and a factor of ~ 2 for electrons upon increasing the light bias from zero to ~ $5 \times 10^{15}$ photons/cm$^2$. To investigate the effects of surface recombination, the probe $\lambda_{ex}$ was changed to 532 nm in the presence of 532 nm light bias (corresponding to absorption depth ~10 μm) [32,35] as shown in **Fig. S7**. Qualitatively similar results were observed for $\lambda_{probe} \approx 532$ nm. For both the cases of 405 nm and 532 nm excitation, beyond a certain high-intensity light bias, the $I_{ph}$ magnitude reduces as shown in **Fig. S6(c)** and **S7(d)** respectively. This reduction is attributed to the effects of light-soaking [36].

The decrease of the $L_d$ at high dc-light bias suggests that the recombination kinetics of the excess carriers is not monomolecular [35]. This was verified by photophysical studies using intensity and time-dependent Photoluminescence (PL) measurements.

**PL measurements**

MAPbBr$_3$ is characterized by absorption, largely in the UV-Vis spectral range (**Fig. S8**). Photo-thermal deflection spectroscopy and light transmission experiments have shown the presence of sub-bandgap states extending to ~ 2.0 eV [32,37]. **Fig. 3(a)** is the plot of PL spectra for 3 eV (405 nm) excitation in reflection and transmission geometry. In reflection mode, the PL spectra exhibit a peak at 2.27 eV (545 nm) with a shoulder at 2.15 eV (575 nm). In transmission mode, the 2.27 eV peak is masked due to self-absorption. The 2.27 eV peak is identified and attributed to the band-to-band transition (free carrier recombination). The 2.15 eV peak has been attributed to different mechanisms viz. radiative recombination from a defect level or to bound exciton recombination[37,38]. The 2.15 eV is observed exclusively in crystals and not in MAPbBr$_3$ thin films (**Fig. S9**). Photo-induced absorption studies at different T of samples with different degrees of crystallinity can be used to attribute the 2.15 eV feature, in the present case, to long-range dipole-dipole interactions of defects in the bulk of single crystals [35,39]. Without loss of generality and for simplicity of modeling and analysis, we identify the 2.15 eV (575 nm) emission as trap mediated radiative emission in the bulk of MSC. This is shown schematically in **Fig. 3(d)**.

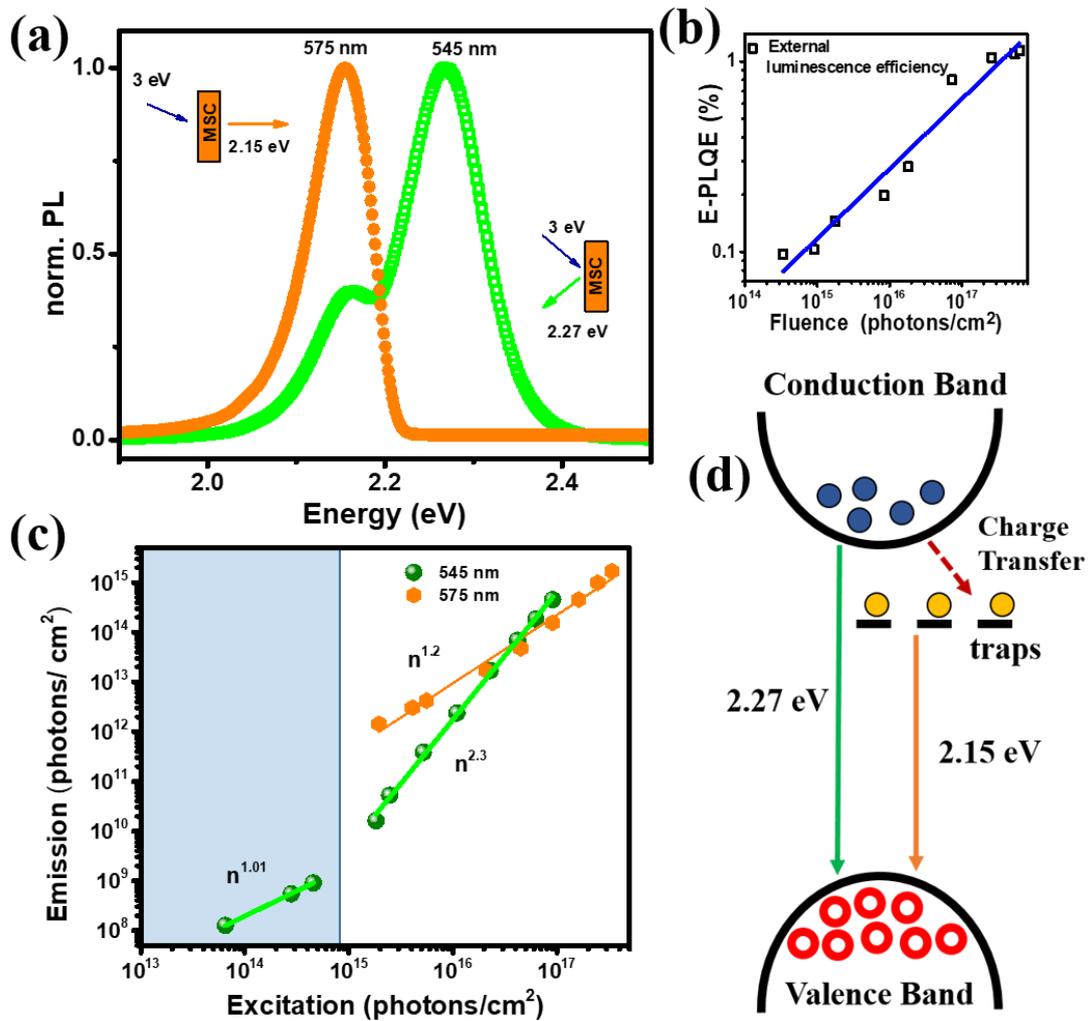

**Fig. 3: (a)** PL data depicts 545 nm and 575 nm peak corresponding to PL measured in the reflection and transmission geometry respectively. **(b)** Plot shows the increase in E-PLQE at higher excitation fluence. **(c)** Emission intensity exhibits linear and quadratic dependence on excitation for 545 nm and linear variation for the 575 nm emission. **(d)** Schematic showing an illustration of band-to-band recombination and trap emission.

**Fig. 3(b)** shows the intensity dependence on External PL Quantum Efficiency (E-PLQE). E-PLQE increases with an increase in the excitation intensity. **Fig. 3(c)** (and also **Fig. S10**) shows that the intensity dependence of emission on the excitation density for the 545 nm and 575 nm emission follows a power law: $<I_{PL}> = <I_{ex}>^x$ and is in agreement with previous observations [37,40]. For the 545 nm peak (2.27eV), $x \approx 1$ at low intensity (blue shaded region) and $\approx 2$ at high intensity. In contrast, the 575nm (2.15 eV) emission peak reveals a linear $x \approx 1$ response throughout. These results can be understood from a simple physical model where the effective excess carrier lifetime ($\tau_{eff}$) is expressed as $\frac{1}{\tau_{eff}} = \frac{1}{\tau_{SRH}} + \frac{1}{\tau_R}$ where, $\tau_{SRH}$ is the trap mediated non-radiative recombination lifetime and $\tau_R$ is the radiative recombination time. The auger recombination component which becomes prominent at high excitation fluence ($> 10^{17}$ photons/cm$^2$) is neglected. $1/\tau_R$ can then be expressed as

$$\frac{1}{\tau_R} = B\,(N_A + \delta n) \; and \; \frac{1}{\tau_{SRH}} = R_{SRH}.N_T \qquad (3)$$

where, $\delta n$ is the excess carrier density, $B$ the radiative constant, and $N_A$ the background doping concentration, $R_{SRH}$ is the non-radiative constant, and $N_T$ is the trap density. The radiative recombination rate is given as:

$$\frac{\delta n}{\tau_R} = A_R \delta n + B\,(\delta n)^2 \qquad (4)$$

where, $A_R = B.N_A$.

At low fluence, the radiative recombination rate increases linearly with the excitation and quadratically with fluence at a higher intensity. This explains the excitation dependence of the 545 nm peak. The 575 nm peak arises due to recombination from a shallow trap. If $\delta n$ is smaller than the trap concentration, the emission rate is linearly dependent on the excitation intensity - in agreement with the observation in **Fig. 3(b)**. This basic model satisfactorily accounts for the PL observations.

**Fig. 4(a)** shows the time-resolved emission spectra upon pulsed 405 nm excitation (19 mW/cm$^2$) on MSC. From the decay trends of the two peaks (**Fig. 4(b)**), it can be observed that 545 nm is characterized by a faster decay lifetime in comparison to the trap emission at 575 nm.

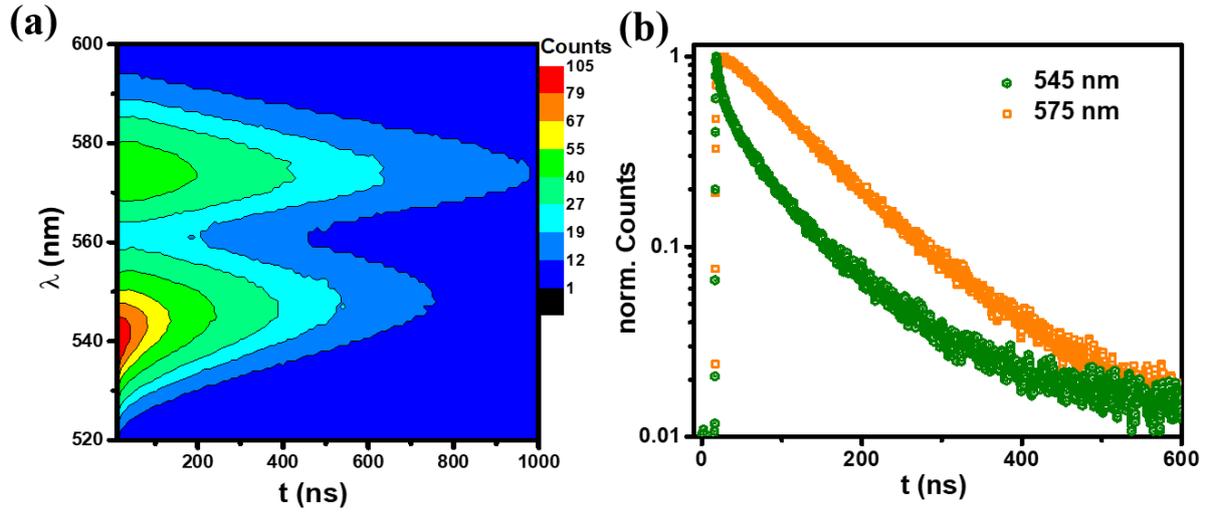

**Fig. 4: (a)** Time-resolved emission spectra on MSC upon pulsed excitation with a 405 nm source. **(b)** TRPL at 545 nm and 575 nm shows a shorter band-recombination lifetime at 545 nm.

The PL-lifetimes were studied by varying the excitation intensity. **Fig. 5(a)** and **5(b)** show the luminescence decay for both the 545 nm and the 575 nm respectively as a function of excitation power. The lifetime of the 545 nm emission decreases with an increase in excitation power. The results are summarized in **Fig. 5(c)**. The lifetime measured in the TRPL corresponds to $\tau_{eff}$ which is given as:

$$\frac{\delta n}{\tau_{eff}} = A\delta n + B\,(\delta n)^2 \qquad (5)$$

where, $A = B \cdot N_A + R_{SRH} \cdot N_T$.

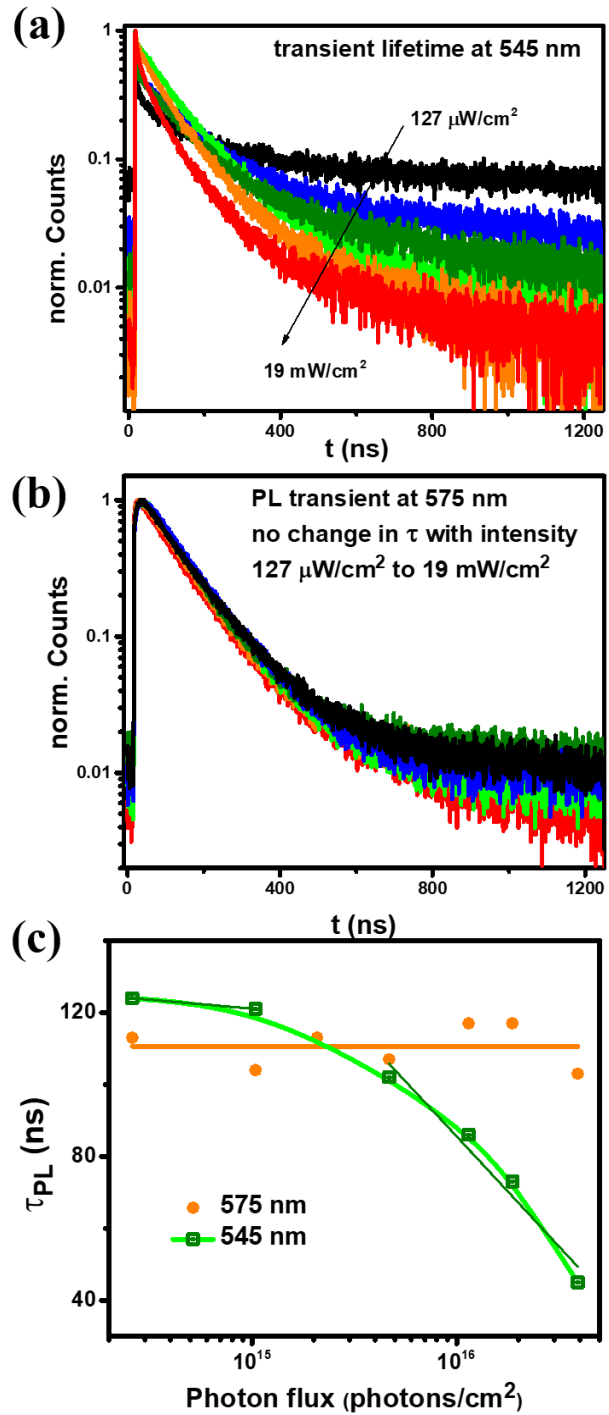

**Fig. 5: (a)** TRPL depicting the decrease in the PL lifetime at high fluence illumination (545 nm) **(b)** TRPL for 575 nm depicts the constancy of lifetime for the range of excitation intensities. **(c)** Compilation of intensity-dependent lifetimes for 545 nm and 575 nm emission. For 545 nm emission, the lifetime is constant in the lower intensity (< $10^{15}$ photons/cm$^2$) and reduces at a higher intensity (> $10^{15}$ photons/cm$^2$). For the 575 nm emission, a uniform lifetime in the entire intensity regime.

The effective PL lifetime corresponding to 545 nm emission, marginally varies up to a fluence of $10^{15}$ photons/cm$^2$ and at higher fluence, decreases as $\tau_R \propto 1/B\delta n$. In contrast, for trap-emission at 575 nm, observed $\tau_{eff}$ is independent of excitation intensity, implying monomolecular emission. Using **Eq. (5)** and the trends observed in **Fig. 5(c)**, the coefficients $A$ and $B$ for 545 nm emission were determined to be $(9.3 \pm 0.2) \times 10^6$ s$^{-1}$ and $(1.4 \pm 0.2) \times 10^{-9}$ cm$^3$s$^{-1}$ respectively (details in Section 12, Supplemental Material). These values are in good agreement with recombination rate constants for direct band-gap semiconductors [32,41]. Since the $E\text{-}PLQE \sim \tau_{eff}/\tau_R$, for a fluence $> 10^{15}$/cm$^2$, $\tau_R$ decreases with increasing fluence. The E-PLQE increases with increasing fluence as shown in **Fig. 3(b)**.

**Discussion**

We now correlate $L_d$ to the PL-lifetime. In the absence of light bias, the tightly-focused probe beam (405 nm, $5\times10^{16}$ photons/cm$^2$) in the SPCM studies generate a high concentration of electron-hole pairs and correspondingly the excess carrier lifetime observed from PL studies (**Fig. 5(c)**) is small (< 28 ns). The excess carriers diffuse outside the illuminated zone and recombine gradually and the corresponding carrier lifetimes at these levels of concentration increase as observed in PL lifetime studies (545 nm, **Fig. 5(c)**). As the modulated probe-carriers diffuse to the electrodes, they are also trapped in deep trapping states effectively reducing the photocurrent. When the probe beam is within a couple of diffusion lengths away from the electrode, an increasing number of carriers (for instance, electrons at the ETL contact) are collected and the holes diffuse to the HTL contact setting up a current in the external circuit. When the carrier concentration reduces sufficiently, the excess carrier lifetime converges to ~ 125 ns, and $L_{diff}$ assumes the low-intensity value. The addition of dc light bias in the SPCM studies results in two effects: 1) DC light bias establishes a new dc equilibrium by populating the deep trapping states. This results in suppression of trapping along the pathway for the carriers generated by the modulated probe beam. The reduced trapping accounts for $I_{ph}$ increase with light bias, explaining features in **Fig. 2(a)** and **Fig. S6(c)**. 2) The magnitude of the dc light bias determines the lateral excess-carrier concentration. The excess carrier lifetime decreases to a limit which is set by the dc-light bias magnitude. This scenario explains the decrease of $L_d$ with increasing light bias.

The light-bias feature elicits the trap-assisted recombination processes[42]. Monomolecular recombination lifetime depends on radiative rate, SRH recombination rate, and trap density as shown in **Eq. (5)**. An estimated trap density of ~ $10^{15}$ traps/cm$^3$ [32,35] significantly influences band recombination and lifetimes. Loss of charge carriers to traps is reduced in the presence of bias light presumably since traps occupancy increases with bias light. Under light bias, a higher fraction of the probe carriers bi-molecularly recombine resulting in shorter $\tau$ and $L_d$.

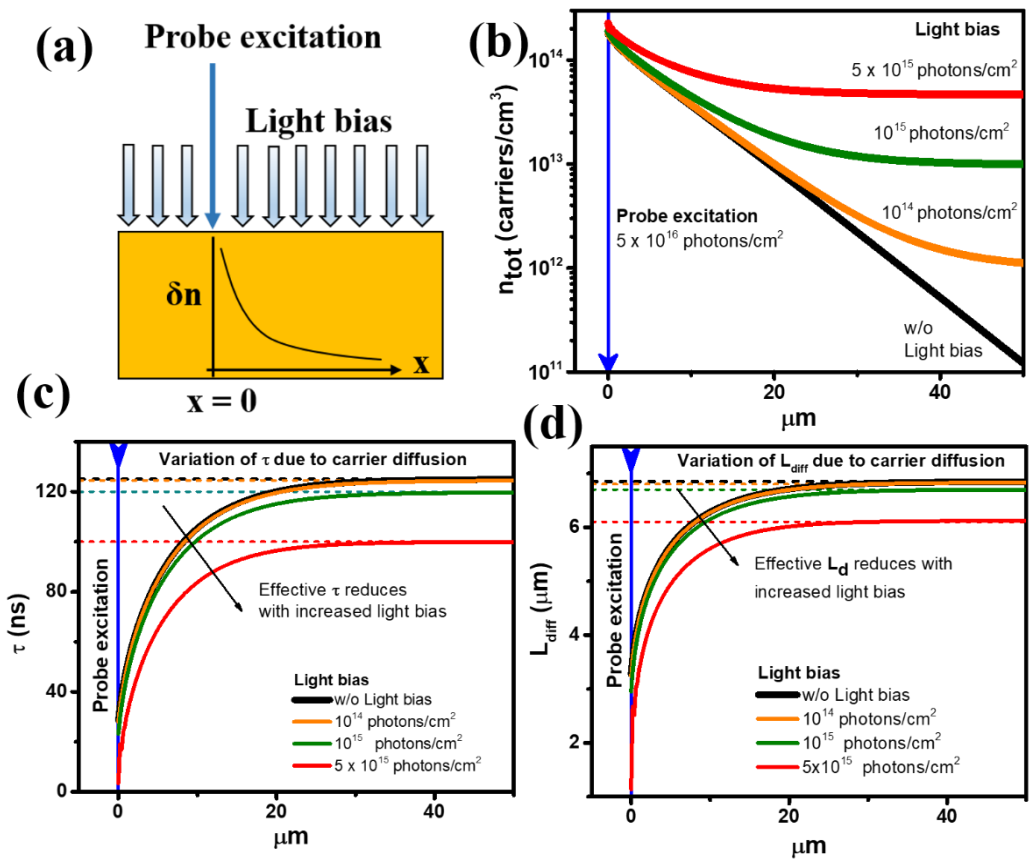

**Fig. 6**: **(a)** The schematic describes the simulation of the SPCM experiment using the finite element method. The probe excitation (at x=0) is assumed to be a point source in addition to uniform dc light bias on MSC sample. The blue line at x=0 in all the plots indicates probe excitation. **(b)** Plot of spatial

carrier decay profile considering 1-D diffusion of probe-carriers. The decay profile changes with different intensities of light bias. **(c)** Shows the excess carrier lifetime at different dc light bias. The excess carrier lifetime depends on the background carrier density (shown in **(b)**). Dashed line indicates lifetime of light bias generated carriers. As excess probe-carriers diffuse in space and recombine, the excess carrier density decreases resulting in an increased excess-carrier lifetime. This results in a spatial dependence of the excess carrier lifetime. **(d)** Shows the effective diffusion length decreases with increasing light bias. Dashed line indicates $L_{diff}$ of light bias generated carriers. Since the lifetime increases as the excess carriers diffuse away from x = 0, the effective diffusion length also increases.

To understand the microscopic carrier diffusion dynamics, simulation of SPCM using the finite element method, incorporating 1-D carrier diffusion was carried out, results of which are presented in **Fig. 6** (simulation details in Section 13, Supplemental Material). A probe excitation is introduced at x = 0 in addition to uniform dc-light bias on the MSC sample, as shown in the schematic **Fig. 6(a)**. The excess carrier concentration at *x* is given as:

$$n_{tot}(x) = \delta n_{probe}(0) \cdot \exp\left(\frac{-x}{L_{diff}}\right) + n_{exc}(L.B) \qquad (6)$$

The first term on the right side of **Eq. (6)** represents the diffusion of probe carriers and the second term corresponds to excess carrier generation due to the light bias. The plot of $n_{tot}(x)$ for different light bias conditions is given in **Fig. 6(b)**. The photogenerated carriers diffuse from x=0 and also simultaneously recombine with lifetime τ which depends on excess carrier concentration (**Fig. 5(c)** and **Fig. S12(b)**). As the excess probe-carriers diffuse away from x=0 the carrier density reduces, increasing τ. However, the effective τ reduces with increasing light bias as shown in **Fig. 6(c)**. A similar trend is observed for the $L_{diff}$, as shown in **Fig. 6(d)**. As the probe carriers diffuse from x=0, the increase in lifetime (**Fig. 5(c)**) results in an increased value of $L_{diff}$. The effective $L_{diff}$ reduces with increased light bias. The simulation results capture the observed trend of decreasing $L_d$ (**Fig. 2(c)**).

The other parameter, apart from lifetime, that determines $L_d$ is the diffusion coefficient. Ščajev et.al has shown that the balancing effects of phonon scattering at low intensity and carrier scattering at high intensity was observed to have marginal effect on $D$[30]. Therefore, the intensity dependent factor that influences diffusion dynamics is determined largely by $\tau$. Photon recycling in perovskites is an additional factor that contributes to the transport length in perovskite based systems and devices [12,43]. In Section 14 of Supplemental Material, we have presented a discussion which shows the simulated excess-carrier density accounting for photon recycling. Considering sizable reabsorbed carriers ($n_\alpha = 0.54$, [44]) and an upper limit on the photon propagation length ($L_\alpha = 1/\alpha \sim 10$ μm for 545 nm), **Fig. S13** shows that the changes in $n_{tot}$, $\tau$ and $L_{diff}$ is marginal upon inclusion of photon recycling.

**Conclusion**

An elegant consistency is observed from the measurements of $I_{ph}(x)$ and PL studies. $L_d$ was observed to decrease as a function of light bias in HOIP single crystals. This observation was correlated with PL lifetime studies which revealed the contribution of trap recombination dynamics in addition to free carrier dynamics. These results provide an insight into the contribution of free carrier and trap emission dynamics to $L_d$. The sizable $L_d$ in systems where carrier transport is trap mediated points to the defect-tolerant capability of HOIP based device structures.

**Acknowledgements**

The authors acknowledge Department of Science and Technology, Government of India, EPSRC-UKRI Global Challenge Research Fund project, SUNRISE (EP/P032591/1) for the financial assistance.

**Conflict of Interest**

The authors declare no conflict of interest.

**References**

[1]     Q. Jiang *et al.*, Nature Photonics **13**, 460 (2019).

[2]     J. Ding *et al.*, Joule **3**, 402 (2019).

[3]     Y. Wang *et al.*, Advanced Materials **29**, 1603995 (2017).

[4]     Y. Wu, X. Li, and H. Zeng, ACS Energy Letters **4**, 673 (2019).

[5]     Y. Jia, R. A. Kerner, A. J. Grede, B. P. Rand, and N. C. Giebink, Nature Photonics **11**, 784 (2017).

[6]     W. S. Yang *et al.*, Science **356**, 1376 (2017).

[7]     N. R. E. Laboratory,  (NREL Golden, CO, 2019).

[8]     Q. Dong, Y. Fang, Y. Shao, P. Mulligan, J. Qiu, L. Cao, and J. Huang, Science **347**, 967 (2015).

[9]     S. D. Stranks, G. E. Eperon, G. Grancini, C. Menelaou, M. J. Alcocer, T. Leijtens, L. M. Herz, A. Petrozza, and H. J. Snaith, Science **342**, 341 (2013).

[10]    Y. Chen *et al.*, Nature communications **7**, 1 (2016).

[11]    T. Etienne, E. Mosconi, and F. De Angelis, The journal of physical chemistry letters **7**, 1638 (2016).

[12]    L. M. Pazos-Outón *et al.*, Science **351**, 1430 (2016).

[13]    C. Wehrenfennig, G. E. Eperon, M. B. Johnston, H. J. Snaith, and L. M. Herz, Advanced materials **26**, 1584 (2014).

[14]    K. X. Steirer, P. Schulz, G. Teeter, V. Stevanovic, M. Yang, K. Zhu, and J. J. Berry, ACS Energy Letters **1**, 360 (2016).

[15]     H. Huang, M. I. Bodnarchuk, S. V. Kershaw, M. V. Kovalenko, and A. L. Rogach, ACS energy letters **2**, 2071 (2017).

[16]     G. W. Adhyaksa, L. W. Veldhuizen, Y. Kuang, S. Brittman, R. E. Schropp, and E. C. Garnett, Chemistry of Materials **28**, 5259 (2016).

[17]     O. E. Semonin *et al.*, The journal of physical chemistry letters **7**, 3510 (2016).

[18]     N. r. F. Montcada, J. M. Marín-Beloqui, W. Cambarau, J. s. Jiménez-López, L. Cabau, K. T. Cho, M. K. Nazeeruddin, and E. Palomares, ACS Energy Letters **2**, 182 (2017).

[19]     M. I. Saidaminov *et al.*, Nature communications **6**, 1 (2015).

[20]     Y. Bi, E. M. Hutter, Y. Fang, Q. Dong, J. Huang, and T. J. Savenije, The journal of physical chemistry letters **7**, 923 (2016).

[21]     D. Shi *et al.*, Science **347**, 519 (2015).

[22]     Z. Xiao, Q. Dong, C. Bi, Y. Shao, Y. Yuan, and J. Huang, Advanced Materials **26**, 6503 (2014).

[23]     Z. Chen *et al.*, Nature communications **8**, 1 (2017).

[24]     Z. Chen, B. Turedi, A. Y. Alsalloum, C. Yang, X. Zheng, I. Gereige, A. AlSaggaf, O. F. Mohammed, and O. M. Bakr, ACS Energy Letters **4**, 1258 (2019).

[25]     Y. Fang, Q. Dong, Y. Shao, Y. Yuan, and J. Huang, Nature Photonics **9**, 679 (2015).

[26]     D. A. Neamen, *Semiconductor physics and devices: basic principles* (New York, NY: McGraw-Hill, 2012).

[27]     H. Scher and E. W. Montroll, Physical Review B **12**, 2455 (1975).

[28]     D. Kabra, S. Shriram, N. Vidhyadhiraja, and K. Narayan, Journal of applied physics **101**, 064510 (2007).

[29]     D. Kabra and K. Narayan, Advanced Materials **19**, 1465 (2007).

[30] P. Ščajev, R. n. Aleksiejūnas, S. Miasojedovas, S. Nargelas, M. Inoue, C. Qin, T. Matsushima, C. Adachi, and S. Juršėnas, The Journal of Physical Chemistry C **121**, 21600 (2017).

[31] G. A. Elbaz, D. B. Straus, O. E. Semonin, T. D. Hull, D. W. Paley, P. Kim, J. S. Owen, C. R. Kagan, and X. Roy, Nano letters **17**, 1727 (2017).

[32] B. Wenger, P. K. Nayak, X. Wen, S. V. Kesava, N. K. Noel, and H. J. Snaith, Nature communications **8**, 1 (2017).

[33] D. W. DeQuilettes *et al.*, Nature communications **7**, 1 (2016).

[34] D. J. Slotcavage, H. I. Karunadasa, and M. D. McGehee, ACS Energy Letters **1**, 1199 (2016).

[35] B. Wu, H. T. Nguyen, Z. Ku, G. Han, D. Giovanni, N. Mathews, H. J. Fan, and T. C. Sum, Advanced Energy Materials **6**, 1600551 (2016).

[36] W. Nie *et al.*, Nature communications **7**, 1 (2016).

[37] H. He *et al.*, Nature communications **7**, 1 (2016).

[38] J. Shi, H. Zhang, Y. Li, J. J. Jasieniak, Y. Li, H. Wu, Y. Luo, D. Li, and Q. Meng, Energy & Environmental Science **11**, 1460 (2018).

[39] G. Grancini *et al.*, Nature photonics **9**, 695 (2015).

[40] T. Schmidt, K. Lischka, and W. Zulehner, Physical Review B **45**, 8989 (1992).

[41] J. S. Manser and P. V. Kamat, Nature Photonics **8**, 737 (2014).

[42] S. D. Stranks, V. M. Burlakov, T. Leijtens, J. M. Ball, A. Goriely, and H. J. Snaith, Physical Review Applied **2**, 034007 (2014).

[43] C. Cho, B. Zhao, G. D. Tainter, J.-Y. Lee, R. H. Friend, D. Di, F. Deschler, and N. C. Greenham, Nature Communications **11**, 1 (2020).

[44] T. Yamada, Y. Yamada, Y. Nakaike, A. Wakamiya, and Y. Kanemitsu, Physical Review Applied **7**, 014001 (2017).

# Supplemental Material

## Impact of Trap Filling on Carrier Diffusion in MAPbBr$_3$ Single Crystals

N. Ganesh, Anaranya Ghorai, Shrreya Krishnamurthy, Suman Banerjee, K.L.Narasimhan, Satishchandra B Ogale, K.S.Narayan*

**List of contents:**

1. Powder XRD on MAPbBr3 single crystals.

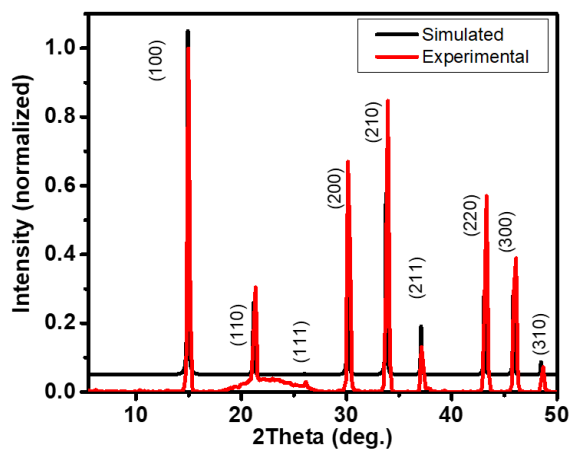

**Fig. S1**: (**a**) Powder XRD on MAPbBr₃ single crystals. Sharp peak indicate long range crystallinity in these samples. The powder XRD data collected on crushed crystals were compared with the simulated pattern obtained from the single crystal XRD analysis. The comparison of the two patterns conclusively affirmed that the phase of the as synthesized single crystals were the same.

2. Knife edge method for spot size determination

In this method of measuring the beam profile, a knife edge is translated across the focused beam in the line of the light incident on the detector. As the knife edge translates the intensity of light on the detector varies as **Fig. S2(a)**. Differential of the intensity gives the beam profile shown in **Fig. S2(b).**

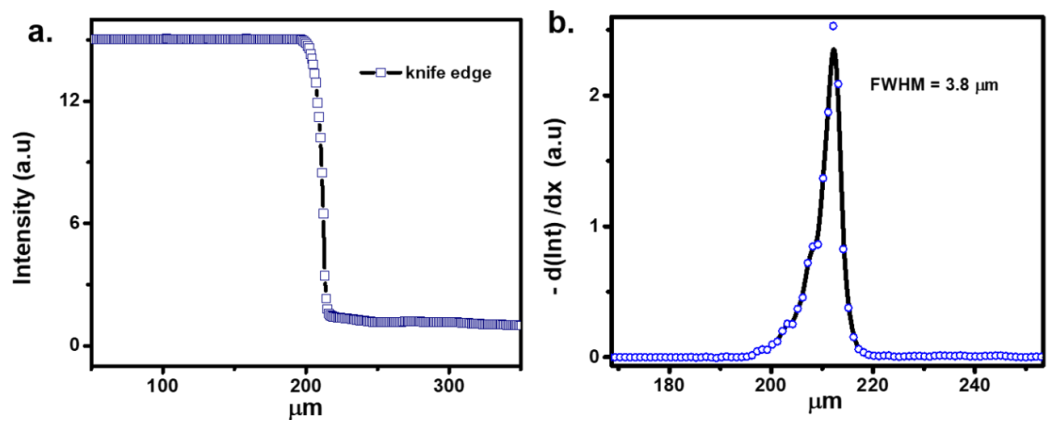

**Fig. S2:** (**a**) The intensity profile as the knife edge is translated across the diameter of the focused beam spot. (**b**) Differential of the intensity profile gives the spot profile and the FWHM of the beam spot.

3. Position dependent Transient Photocurrent on single crystal device

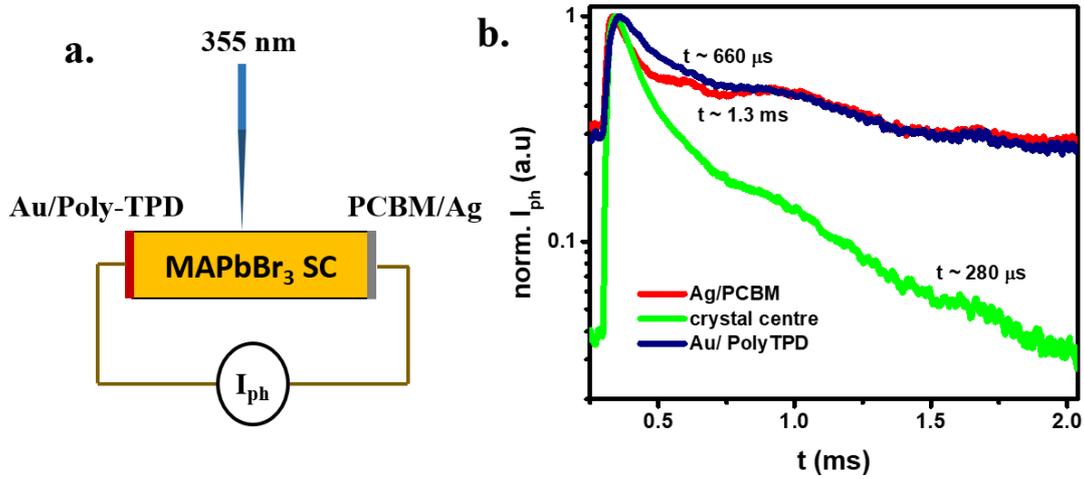

**Fig. S3: (a)** Schematic of the position dependent TPC on the MSC device. The 355 nm pulsed laser, 1 ns pulse, 60 nJ/pulse was for excitation close to the interface and at the center of the crystal. **(b)** The TPC decay time shows longer decay time for excitation close to electrode when compared to excitation in the central region far from the electrodes. The long decay times reflect on the transit time of the un-extracted counter carrier.

4. Invariance of $L_d$ with probe beam excitation intensity.

We have varied the probe intensity by a factor of ~ 3, since the probe intensity is within the bimolecular regime of excitation, corresponding to which the lifetime variation is expected to be drastic. We also note that, at very low probe beam intensity (< 1 mW/cm$^2$), the $I_{ph}$ magnitude reduces to background current levels, resulting in the absence of the observed $I_{ph}(x)$ features.

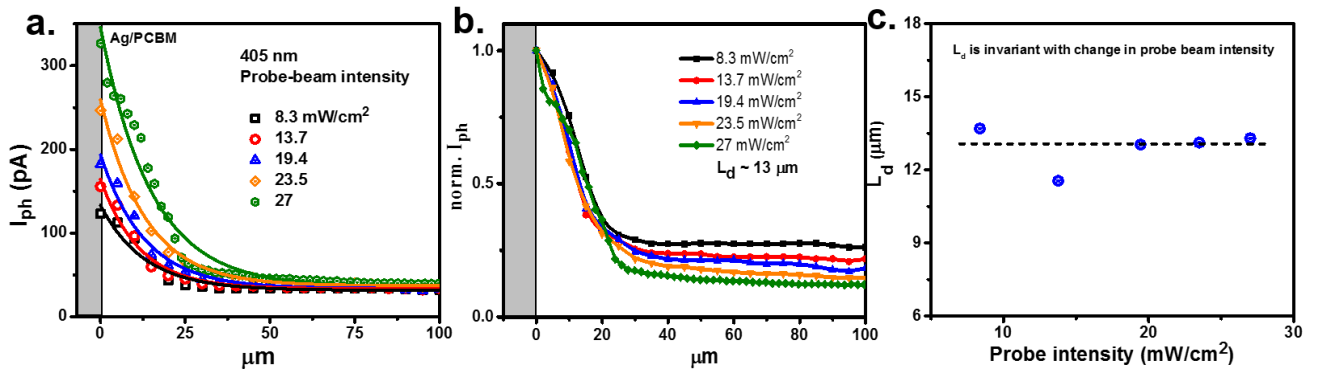

**Fig. S4: (a)** Plot of $I_{ph}(x)$ with variation in the 405 nm probe beam intensity in the absence of light bias. Solid lines indicate single exponential fits to obtain $L_d$. **(b)** Normalized plot of $I_{ph}(x)$ for different probe

beam intensities with SPCM across the Ag/PCBM-MSC interface. **(c)** Plot of $L_d$ with respect to probe beam intensity indicates that $L_d$ remains invariant with change in probe-beam intensity.

5. Effect of Recovery time on $I_{ph}(x)$ scan

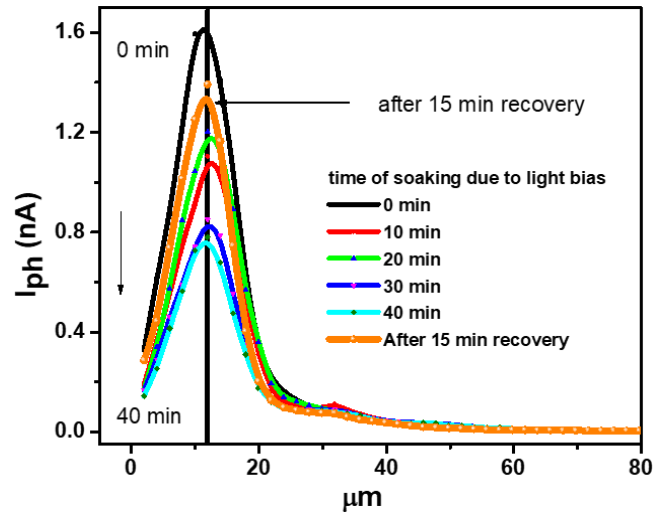

**Fig. S5:** The effect of recovery time on MSC devices were seen on the $I_{ph}(x)$ scan across the MSC-electrode (PolyTPD-Au) interface. Each of the experiments were carried out with 405 nm probe beam and 390 nm light bias at ~ 2 mW/cm². Each of the $I_{ph}$ scans is performed after exposure to light bias. At the end of 40 min, the system was in recovery condition: in the dark at short circuit condition with a positive pressure of inert atmosphere for 15 min.

6. Additional plots for 390 nm light bias and maximum I$_{ph}$ under different light bias intensity.

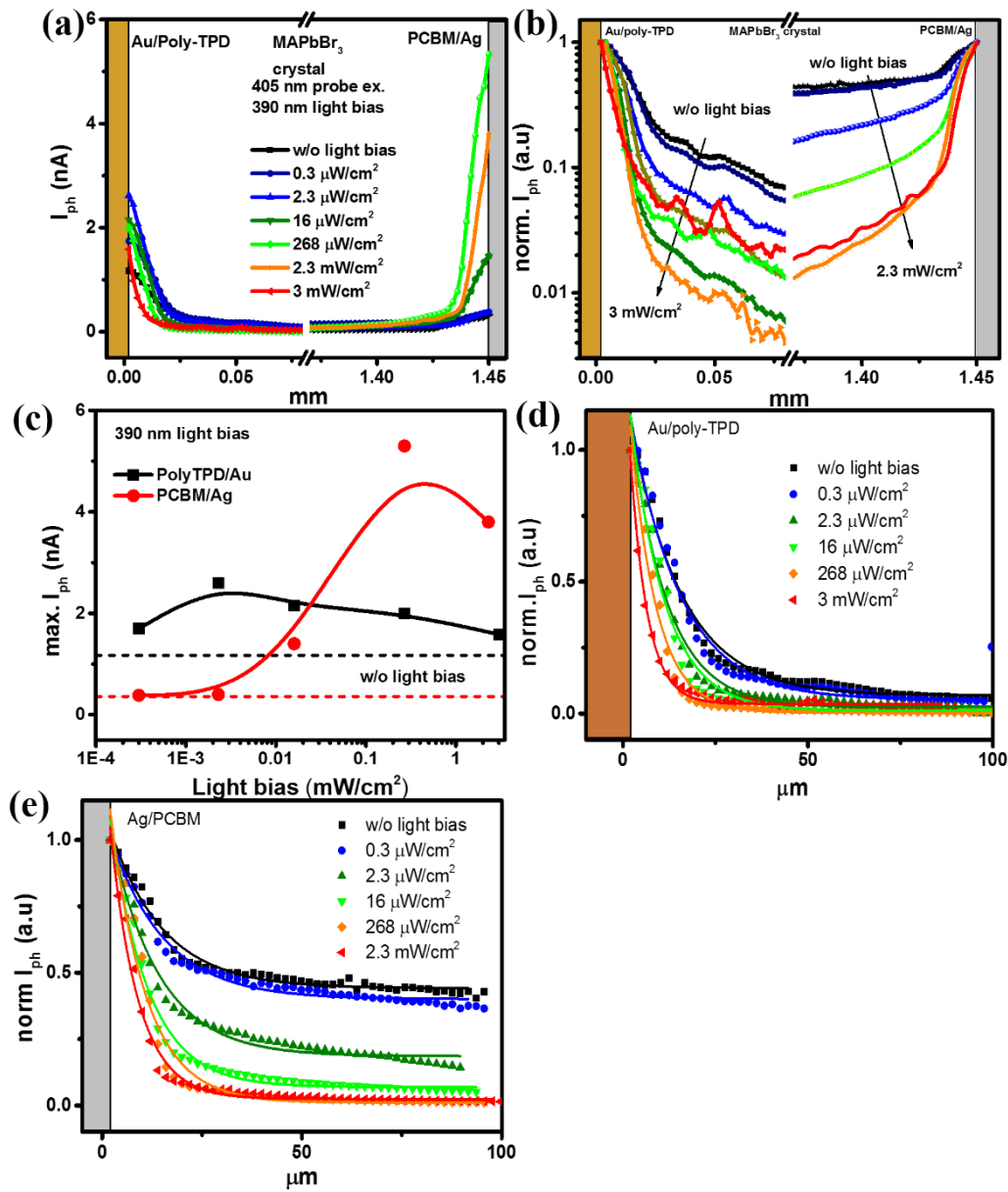

**Fig. S6**: **(a)** Plot of I$_{ph}$(x) profiles obtained from SPCM experiments using 405 nm probe excitation and 390 nm light bias. The I$_{ph}$(x) variation is seen for different intensities of light bias. **(b)** Normalized I$_{ph}$(x) plot in semi-log scale showing sharp I$_{ph}$ decay for the case of higher light bias. **(c)** Plot of variation of max. I$_{ph}$ close to the electrode as a function of light bias intensity for 390 nm. The dashed horizontal lines correspond to max. I$_{ph}$ without the light bias. **(d)** Normalized I$_{ph}$ plots (with electrode interface considered as the origin) with mono-exponential decay fits for SPCM scans across the Au/PolyTPD-MSC **(e)** Ag/PCBM-MSC interface respectively.

7. SPCM scans with 532 nm probe excitation and light bias.

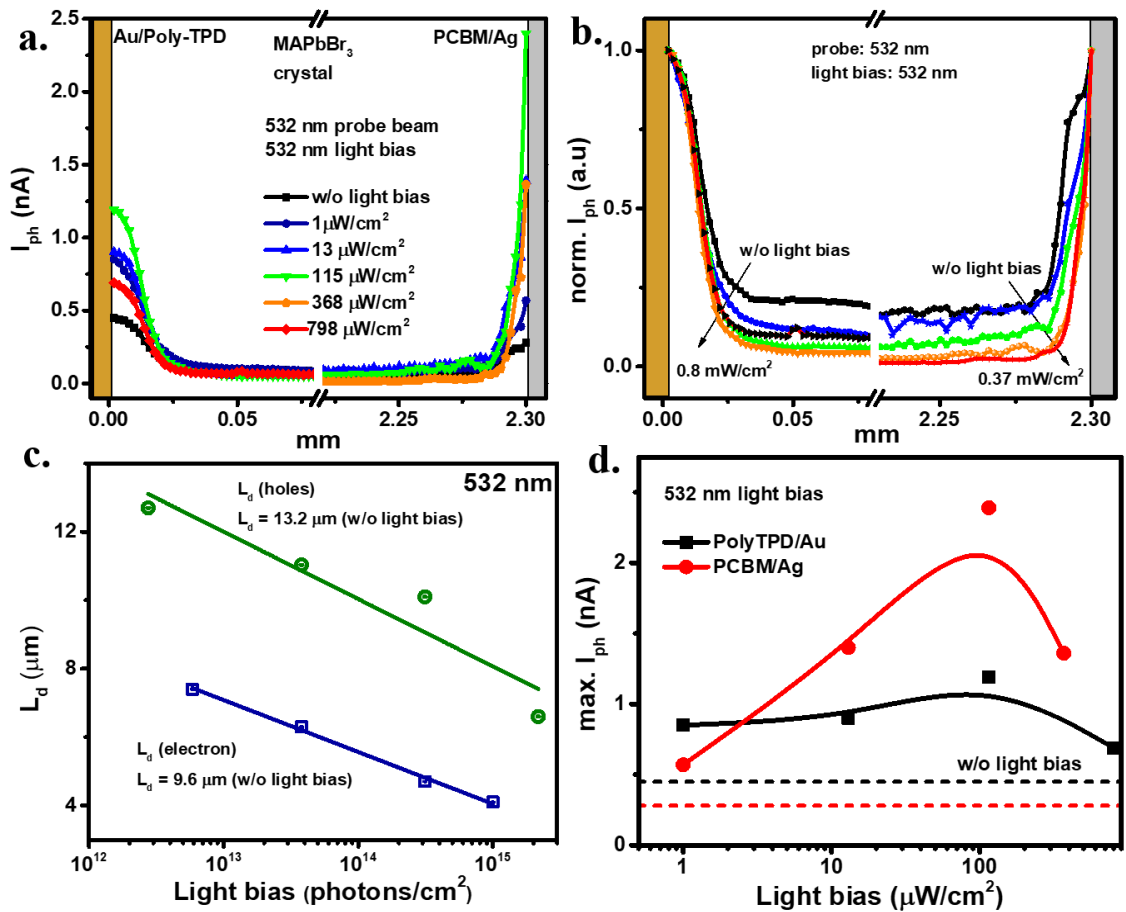

**Fig. S7**: (**a**) $I_{ph}(x)$ variation on either ends of the MSC device with 532 nm probe beam and 532 nm light bias. (**b**) Normalized $I_{ph}(x)$ profiles indicating sharp decay at higher light bias. (**c**) $L_d$ extracted with exponential fitting of the $I_{ph}(x)$ decreases as a function of light bias-photon flux. (**d**) Plot of variation of max. $I_{ph}$ close to the electrode as a function of light bias intensity for 532 nm. The dashed horizontal lines correspond to max. $I_{ph}$ without the light bias.

8. Absorption spectra of MAPbBr3 single crystals

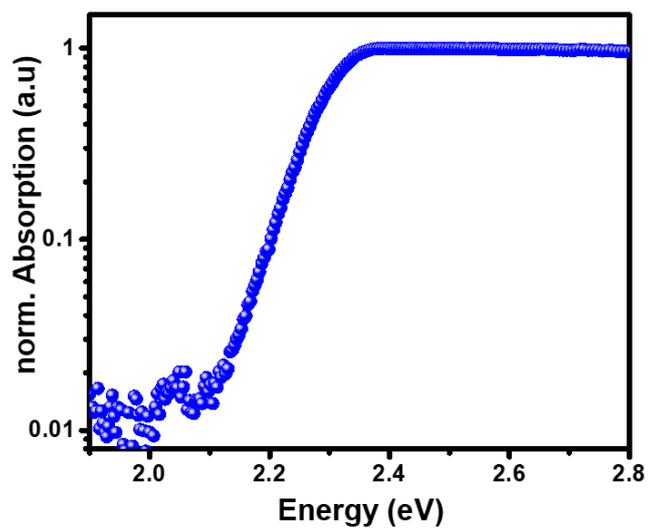

**Fig. S8**: Normalized absorption spectra of MAPbBr$_3$ single crystal.

9. PL of thin films and single crystals.

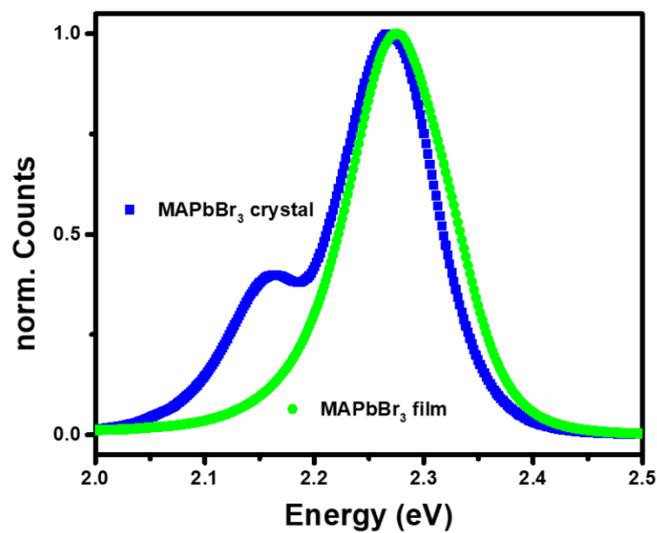

**Fig. S9:** PL spectrum of MAPbBr3 thin film and single crystals. The emission peak at 2.15 eV (575 nm) corresponding to bulk trap emission is absent in thin films.

10. Intensity dependent PL variation.

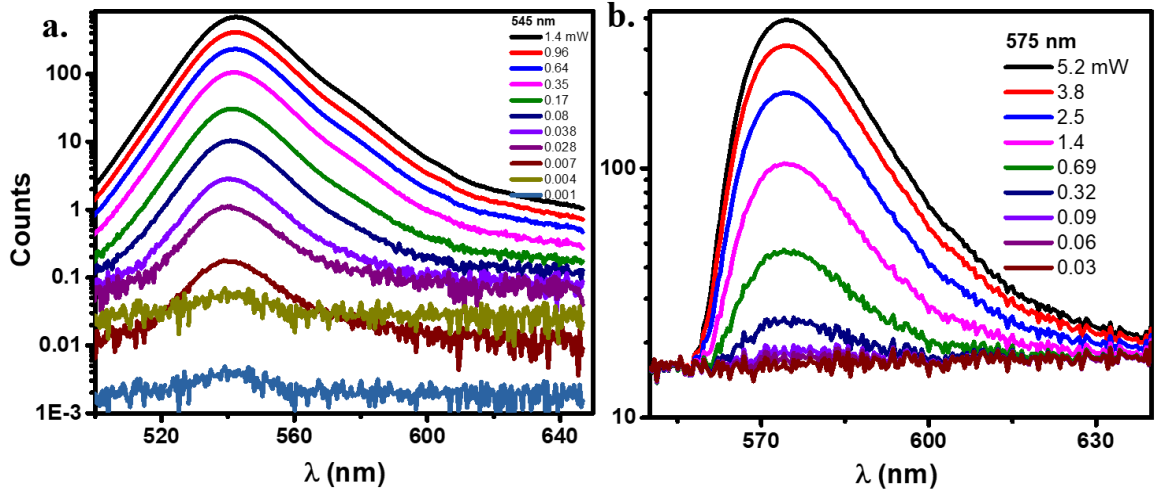

**Fig. S10:** Intensity dependent PL spectrum centered at **(a)** 545 nm, with PL measured in the reflection geometry and **(b)** 575nm, with PL measured in the transmission geometry on single crystals.

11. Drift diffusion formalism for $I_{ph}(x)$ from SPCM

To extract the diffusion length of the carrier from the SPCM, we need to establish the relation of the diffusion length on the observed $I_{ph}$ decay. The drift diffusion equation for a single type of excess carrier is given as:

$$\frac{\partial \delta n}{\partial t} = D \frac{\partial^2 \delta n}{\partial x^2} + \mu E \frac{\partial \delta n}{\partial x} + G - \frac{\delta n}{\tau} \qquad \text{(S1)}$$

The generation rate G is given by delta function, $G = G_0 \, \delta(x-x_0)$ where $x=x_0$ is the region of illumination. Under the conditions of steady state illumination ($d\delta n/dt = 0$), negligibly small electric field ($E = 0$) and $x \neq x_0$ the diffusion equation can be written as:

$$D \frac{\partial^2 \delta n}{\partial x^2} = \frac{\delta n}{\tau}$$

With the diffusion length defined as $L_{diff} = \sqrt{(D\tau)}$, by solving the above equation we obtain,

$$\delta n(x) = \delta n_0 \exp\left(-\frac{x}{L_{diff}}\right)$$

For x > 0, is the distance between the collection electrode and the point of generation. $\delta n_o$ is the generation density at the point excitation.

The obtained $I_{ph}$ as a result of diffusion is given as:

$$I_{ph}(x) = -eD \frac{\partial \delta n}{\partial x}$$

Since $L_{diff}$ is not a constant and has a dependence on $\delta n(x)$, the $I_{ph}(x)$ can be expressed as:

$$\boldsymbol{I_{ph}(x) = I_0 \exp(-\frac{x}{L_d})} \qquad \text{(S2)}$$

Where $L_d$ is the effective diffusion length and $L_d \propto L_{diff}(x) = (kT\mu\tau/e)^{1/2}$, here $\tau$ depends on excess carrier density.

12. Calculation of *A* and *B* coefficients.

The dependence of A and B on the recombination rate is given in **Eq. (5)**. We observe that at low intensities, $\tau = 1/A$ at low intensity and at high intensities, $\tau = 1/B\delta n$.

No of carriers generated is given as $= \alpha G'$, where *G'* corresponds to generation rate per pulse. From the integrated generation rate *G* indicated in **Fig. 5(c)**, $G' = G/(rep.\ rate)$, where *rep. rate* is the repetition rate of the pulsed illumination (800 KHz).

With the $\tau$ values given in **Fig. 5(c)**, the recombination rate is given as: $\delta n/\tau$. To determine *A* and *B*, **Eq. (5)** was fit to a parabolic equation as shown in **Fig. S11**. The values of *A* and *B* were determined to be $(9.3 \pm 0.27) \times 10^6\ s^{-1}$ and $(1.4 \pm 0.21) \times 10^{-9}\ cm^3 s^{-1}$ respectively.

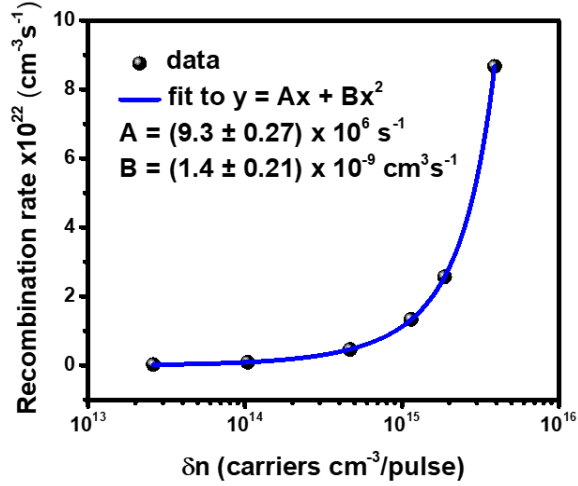

**Fig. S11**: Plot of recombination rate vs excess carrier density. The plot was fit to a quadratic equation to obtain A and B coefficients.

13. Simulation of SPCM with light bias.

We simulated the SPCM experiment with light bias conditions and correlated with PL lifetime values to obtain the $I_{ph}(x)$ profile using the finite element method. We computed at each point the semiconductor characteristics such as: $n_{tot}$, $\tau$, $L_{diff}$ and $I_{ph}$. For simplicity, we have assumed point illumination of the probe excitation and considered 1-D carrier diffusion.

**Fig. S12(a)** shows the schematic of the sample and illumination profile. The probe generation (405 nm, 30 mW/cm$^2$) is at x=0 in addition to dc-uniform light bias on the MSC sample. The probe point excitation consisted of G (0) = 5 x 10$^{16}$ photons/cm$^2$. Upon probe photo-excitation, the generation of excess carrier concentration $\delta n_{probe}$ is given as:

$$\delta n_{probe} = \alpha G_0 \tau$$

$\alpha$ = 8 x 10$^4$ cm$^{-1}$ is the absorption coefficient [1]. The lifetime for a given carrier density is obtained from the linear dependence of $1/\tau$ on excess carrier density, **Fig. S12(b)**. This plot was obtained from lifetime dependence on generation density parameter shown in **Fig. 5(c)**.

The excess carrier concentration, $n_{tot}$, due to both probe carrier diffusion and light bias generation, at each point is:

$$n_{tot}(x) = \delta n_{probe}(0) \exp\left(\frac{-x}{L_{diff}}\right) + n_{exc}(L.B) \quad \text{(S3)}$$

The first term on the RHS refers to probe carrier diffusion with the diffusion length $L_{diff} = (kT\mu\tau/e)^{1/2}$, where $\mu = 115$ cm$^2$/Vs is the charge carrier mobility [2]. The second term corresponds to the excess carrier generated due to light bias such that, $n_{exc}(L.B) = \alpha\tau G_{L.B}$, where $G_{L.B}$ is the generation density of the light bias.

**Fig. S12(c)** shows the plot of excess-carrier decay profile considering **Eq. (S3)**. It is observed that the decay profile changes with the intensity of light bias. As excess probe-carriers diffuse in space and recombine, the excess carrier density decreases resulting in an increased excess-carrier lifetime. This results in a spatial dependence of the excess carrier lifetime. This trend is observed in **Fig. S12(d)**. Dashed lines in figure indicate the carrier lifetimes of the background-light bias carriers. By comparing the lifetime variation, it can be seen that the effective lifetime reduces with increased light bias. As a consequence of lifetime variation, the $L_{diff}$ parameter show similar trend in **Fig. S12(e)**. Dashed line indicates the $L_{diff}$ for background light-bias carriers. As the probe carriers diffuse from x=0, the increase in lifetime results in increased value of $L_{diff}$. The effective $L_{diff}$ reduces with increased light bias. Effect of light bias-modified $L_{diff}$ on probe-excess carriers is shown in the **Fig. S12(f)**. The decay length is seen to reduce at high intensity of light bias. These above mentioned simulation results qualitatively explains the observation of decreasing $L_d$ shown in **Fig. 3(d)**.

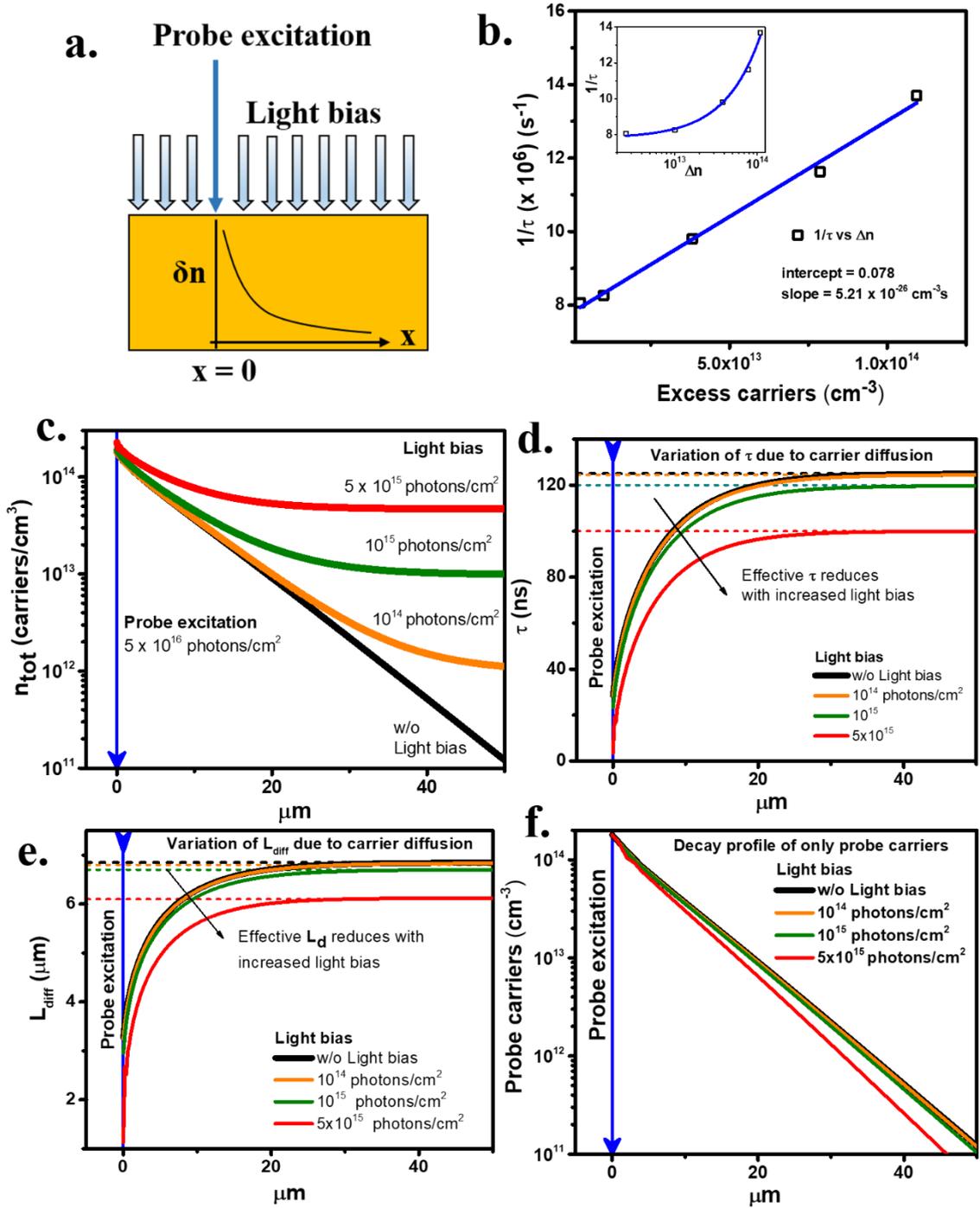

**Fig. S12**: **(a)** The schematic describes the simulation of SPCM experiment using finite element method. The probe excitation (at x=0) is assumed to be a point source in addition to uniform dc light bias on MSC sample. Blue line at x=0 in all the plots indicates probe excitation. **(b)** Linear fit to the plot of PL $1/\tau$ vs excess carrier density. **Inset** shows the plot in semi-log scale **(c)** Plot of spatial carrier decay profile considering 1-D diffusion of probe-carriers. The decay profile changes with different intensities of light bias. **(d)** Shows the excess carrier lifetime at different dc light bias. The excess carrier lifetime depends on the background carrier density (shown in **b**). Dashed line indicates lifetime of light bias generated carriers. As excess probe-carriers diffuse in space and recombine, the excess carrier

density decreases resulting in an increased excess-carrier lifetime. This results in a spatial dependence of the excess carrier lifetime. **(e)** Shows the effective diffusion length decreases with increasing light bias. Dashed line indicates $L_{diff}$ of light bias generated carriers. Since the excess carrier lifetime increases as the excess carriers diffuse away from x = 0, the effective diffusion length also increases. **(f)** Effect on probe carrier diffusion due to change in the $L_{diff}$ (seen in **e**) due to light bias. The decay length is smaller in the case of high intensity light bias.

14. Effect of Photon recycling on excess carrier density, lifetime and diffusion length.

The excess carrier diffusion and photon propagation upon point illumination at x=0 was simulated with finite element method using the coupled equations:

$$n(x) = n_0 exp\left(-\frac{x}{L_{diff}}\right) + \gamma_{pp}(x).n_\alpha.\tau \quad \textbf{(S4)}, \text{ and}$$

$$\gamma(x) = \frac{n(x).\varphi_{PL}}{\tau} + \gamma_{pp}(x)(1-n_\alpha) \quad \textbf{(S5)}$$

Where n(x) and ɡ(x) is the excess carrier density and photon density at point x respectively. $n_\alpha$ is the ratio of reabsorbed photons of the total emitted photons. We consider the value = 0.54 as given in the reference[3]. $f_{PL}$ is the internal luminescence efficiency of the sample and is given by Yamada et. al as[3]:

$$\varphi_{PL} = \frac{\tau_\infty - \tau_0}{\tau_\infty - n_\alpha.\tau_0} \quad \textbf{(S6)}$$

Where $\tau_\infty$ can be attributed to the monomolecular lifetime (~ 125 ns) and $\tau_0$ is the lifetime under consideration for a given carrier density. Considering probe illumination at x=0, the excess density is highest at x=0 and then decays further away (shown in black solid line in **Fig.**

**S12(c))**. Since the lifetime $\tau_0$ is small for high carrier density, the $f_{PL}$ value is higher. **Fig. S13(a)** shows the plot of $f_{PL}$ as a function of distance away from the point of excitation. In proportional to the carrier density, $f_{PL}$ reduces away from the x=0. The photon recycling (PR) efficiency[3] given as $h_{PR} = n_\alpha \cdot f_{PL}$ follows the same trend as $f_{PL}$ (shown in **Fig. S13(a)**)

$g_{PP}(x)$ is the photon density as a result of photon propagation from a previous emissive event. It can be written as:

$$\gamma_{PP}(x) = \gamma(x_{n-1}) \cdot \exp\left(\frac{-x}{L_\alpha}\right) \cdot f(\theta)$$

Where $g(x_{n-1})$ is the photon density from the previous iterative determination. $L\alpha = 1/\alpha$ is the photon propagation length. We consider an upper limit of $L\alpha = 10$ μm for the case of 545 nm[1]. $f(\theta)$ is an additional angular function that indicates the fraction of isotropically emitted photons towards 1-D carrier diffusion. Here $\theta = \tan^{-1}(L_\alpha/x)$ is the angle subtended by the emission point, which is x distance away from the collection electrode. In finite element method given the small inter-iterative distance (~ 0.1 μm), i.e., $x \ll L_\alpha$, $f(\theta) \sim 0.5$. This means, in a small distance, half the emitted photons propagate in one direction and the rest in the opposite direction. Additionally, $g(0) = \alpha G_0$ and $n_0 = \alpha G_0 \tau$.

The simulated plot of excess carrier density with and without photon recycling is shown in **Fig. S13(b)**. It can be observed that the change in the carrier density is marginal, and close to the illumination point. Similarly, as shown in **Fig. S13(c)** and **Fig. S13(d)** the change in lifetime and $L_{diff}$ values is marginal. However, we note that the effective lifetime and associated $L_{diff}$ reduces as a consequence of photon recycling, an observation reported preciously.

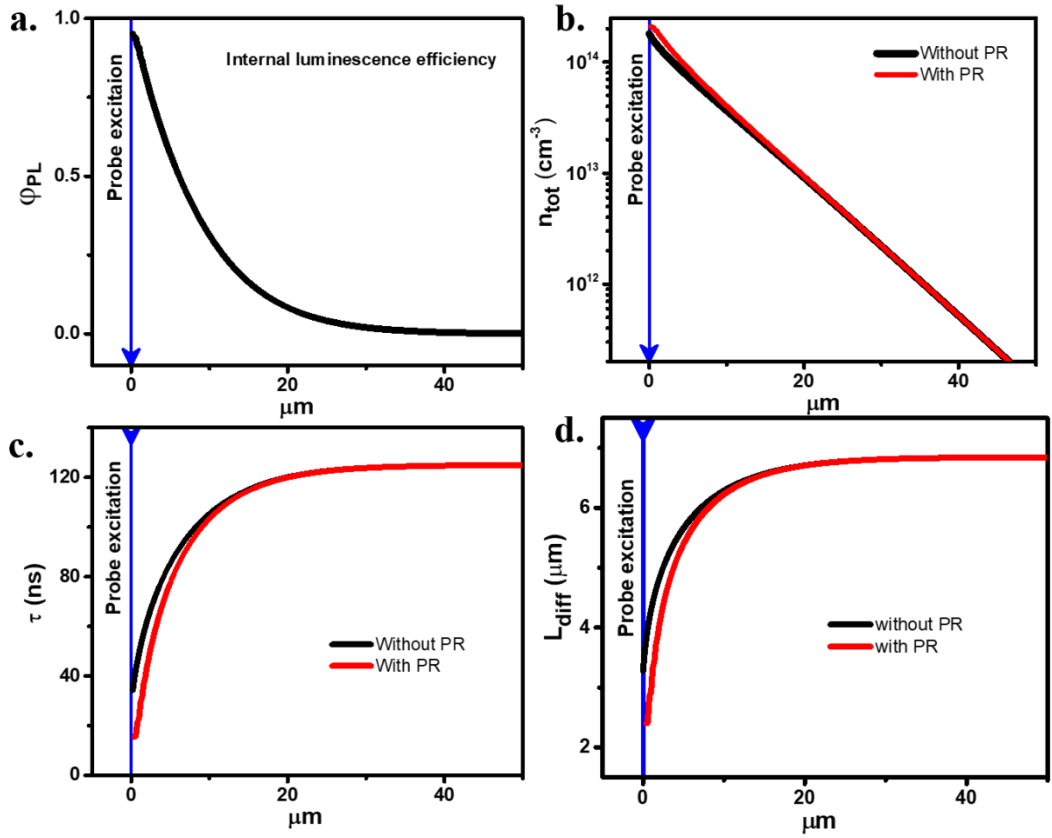

**Fig. S13:** **(a)** Plot of $f_{PL}$ (internal luminescence efficiency) which depends on the background carrier density. Blue vertical lines in all the plots indicate point of excitation. As the carriers diffuse away from x=0, the carrier density reduces resulting in reduction of lifetime (shown in **Fig. S12(c)**). From **Eq. (S6)**, $f_{PL}$ is high when carrier lifetime is smaller. **(b)** Plot of excess carrier density profile shows marginal increase considering PR. Consequently, due to the relative increase in the excess carrier density, the lifetime and diffusion length parameter show a marginal reduction as shown in **(c)** and **(d)** respectively.

15. **Table S1**: $L_d$ values from the SPCM experiment. The values were obtained upon fitting Eq. (2) to the $I_{ph}(x)$ profiles in **Fig. 3(b)**.

| Light bias (photons/cm$^2$) | $L_d$ (μm) Ag/PCBM | $L_d$ (μm) Au/Poly-TPD |
|---|---|---|
| w/o light bias | 13.3 ± 0.6 | 13.8 ± 0.5 |
| 6.12 x 10$^{11}$ | 13.0 ± 0.6 | 11.9 ± 0.4 |
| 4.59 x 10$^{12}$ | 12.1 ± 0.5 | 9.9 ± 0.3 |
| 3.27 x 10$^{13}$ | 9.4 ± 0.3 | 9.3 ± 0.4 |
| 5.41 x 10$^{14}$ | 8.9 ± 0.5 | 7.0 ± 0.2 |
| 4.59 x 10$^{15}$ | 6.9 ± 0.2 | |
| 6.12 x 10$^{15}$ | | 4.6 ± 0.1 |

16. Experimental Section

*Preparation of single crystals and crystal device:*

The crystals were prepared by inverse temperature crystallization technique as reported earlier by Bakr et.al [4]. A 5ml vial was used for the crystal growth containing 1M solution of PbBr$_2$ (98%, Aldrich, mol.wt. 367gm/mol) and MABr (Sigma Aldrich, mol.wt. 111.97gm/mol), using DMF (anhydrous 98.9%, Sigma Aldrich, mol.wt: 73.09gm/mol) as the solvent. The ratio of PbBr$_2$: MABr used was 1:1. The solution was kept in an oil bath wherein the temperature was gradually increased till inverse temperature of crystallization (~80°C) was reached. The seeds (MAPbBr$_3$) formed were further transferred to a new solution for larger crystal growth and the process was repeated to achieve crystals of desired size.

The crystal device was prepared by drop-casting PC$_{71}$BM (Lumtec Corp.) and Poly-TPD (Lumtec Corp.) on opposite sides of the bulk single crystal. 3 mg/ml solution of PCBM and Poly-TPD in anhyd. Chlorobenzene (Sigma Aldrich) were drop-cast in a controlled manner with a micro-pipette. After each drop-cast, the layer was left to anneal by placing the coated crystal on hot-plate at 60° C for 30 min. This was followed by coating Ag (~ 150 nm) on the PCBM layer and Au (~ 40 nm) on the Poly-TPD layer by thermal evaporation.

*Absorption on single crystals.*

The steady state absorption was collected on the MAPbBr$_3$ single crystals using UV-Vis Helios Ultrafast Systems spectrophotometer and integrating sphere. The single crystal was mounted between grooved quartz plates and the spectrum was recorded in reflectance mode.

*PL on single crystals*

The PL on the single crystals were carried out both in the reflection as well as the transmission geometry. A 405 nm diode laser was used to excite the single crystal. In the reflection geometry, using a lens, the light, after passing through a 450 nm long pass filter was collected into the fiber coupled Mini Spectrometer (Hamamatsu, C10083CA).

In the transmission geometry, maintaining the angled incidence of the excitation light, PL was measured with the help of suitable collection optics.

*PL Quantum Efficiency:*

PL measurements were carried out with sample placed in the integrating sphere to determine Ext. PLQE. A 405 nm laser with calibrated intensity was used as the illumination source and fiber coupled mini-spectrometer (Hamamatsu, C10083CA) was used to collect the excitation and emission spectrum. Measurements were carried out with the both the reference slide (glass) and the bulk single crystal placed on the glass slide. Suitable optics was utilized to collimate the beam with spot size smaller than the crystal size. The obtained spectrum was normalized with the responsivity of the detector and the integrated intensity of the excitation ($I_{exc}$) and emission range ($I_{em}$) was used to determine:

$$External\ PLQE\ =\frac{I_{em}(\text{crystal}) - I_{em}(\text{ref})}{I_{exc}(\text{ref}) - I_{exc}(\text{crystal})}$$

*Scanning Photocurrent Microscopy (SPCM)*:

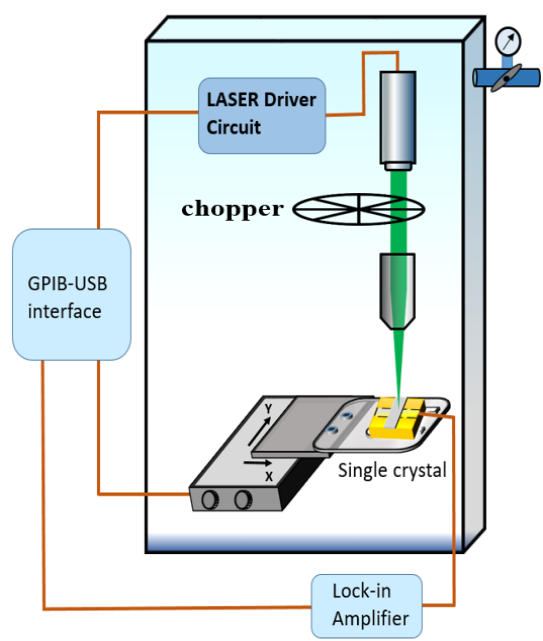

**Fig. S14**: Schematic of SPCM experimental setup

The schematic explaining the SPCM scanning setup is given in the above Figure. A modulated laser light (119 Hz) is focused using a microscope objective (50x, 0.7 NA) on the MSC device. As the device translates, the short circuit $I_{ph}$ is measured using an amplifier (SRS 830), locked-in to the modulating frequency of the laser driver. Additionally, the light bias is introduced by an LED ring-illumination to have uniform constant illumination. The intensity of the light bias was controlled by varying the current sourced to the LEDs. The intensity of the light bias were verified using a calibrated photodiode. The experiments were carried out using a GPIB-USB interfaced program.

To consider the effects of light bias on the probe beam spot size, we model the Gaussian probe beam and check for the change in the spot-size in the presence of light bias. Fig. S15 shows the comparison of the probe-beam with and without the introduction of light bias. We observe that the beam profile and the FWHM remain mostly unaffected by the introduction of light bias.

We, therefore, conclude that the reduction of $L_d$ with the light bias is due to the effect of light bias on the probe beam.

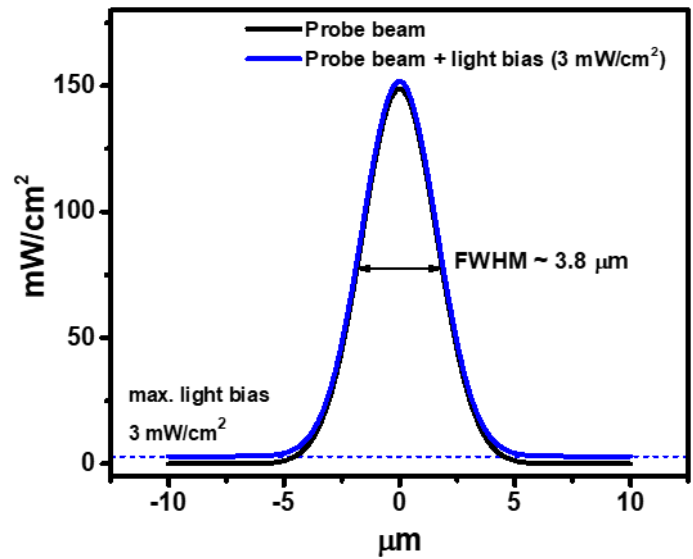

**Fig. S15**: Simulated probe beam profiles (FWHM = 3.8 µm and average power of ~ 30 mW/cm$^2$) with and without light bias. The effect of light bias has a marginal effect on the beam spot size.

*Time resolved Photoluminescence (TRPL)*

Time Resolved Photoluminescence measurements were carried out using a custom built Time Correlated Single Photon Counting (TCSPC) instrument with a temporal resolution of ~250 ps. A Picoquant LDH-P-C-405M laser source with centre wavelength 405 nm and pulse width 90 ps was driven a Picoquant PDL 800-B laser diode driver to excite the crystal. A Tektronix AFG1024 arbitrary Function Generator was used to control the repitation rate and the laser was driven at 800KHz. The emission profile was recorded using a single photon counter (Picoquant PMA-C-192-M) coupled to Zolix Omni 1300 monochromator. Time tagging of the data was done using Picoquant TimeHarp 260 data acquisition card. The time resolved decay profiles were obtained for the wavelength range 500 nm to 700 nm. The time resolved emission spectrum (TRES) was derived from the decay profiles using MATLAB.

**References.**

[1] B. Wenger, P. K. Nayak, X. Wen, S. V. Kesava, N. K. Noel, and H. J. Snaith, Nature communications **8**, 1 (2017).

[2] D. Shi *et al.*, Science **347**, 519 (2015).

[3] T. Yamada, Y. Yamada, Y. Nakaike, A. Wakamiya, and Y. Kanemitsu, Physical Review Applied **7**, 014001 (2017).

[4] M. I. Saidaminov *et al.*, Nature communications **6**, 1 (2015).